\documentclass[english]{article}
\usepackage[T1]{fontenc}
\usepackage[latin1]{inputenc}
\usepackage{geometry}
\usepackage{color}
\usepackage{graphicx}

\makeatletter

\usepackage{babel}
\makeatother
\begin{document}
\begin{center}
{\LARGE Reduction of Quantum Phase Fluctuations in Intermediate States}
\par\end{center}{\LARGE \par}

\begin{center}
{\large Amit Verma}%
\footnote{amit.verma@jiit.ac.in%
} {\large and Anirban Pathak}%
\footnote{anirban.pathak@jiit.ac.in%
} {\large }
\par\end{center}{\large \par}

\begin{center}
Jaypee Institute of Information Technology University, A-10, Sector-62,
Noida, UP-201307
\par\end{center}

\begin{center}
\textbf{INDIA}
\par\end{center}

\begin{abstract}
Recently we have shown that the reduction of the Carruthers-Nieto
symmetric quantum phase fluctuation parameter $(U)$ with respect to
its coherent state value corresponds to an antibunched state, but
the converse is not true. Consequently reduction of $U$ is a
stronger criterion of nonclassicality than the lowest order
antibunching. Here we have studied the possibilities of reduction of
$U$ in intermediate states by using the Barnett Pegg formalism. We
have shown that the reduction of phase fluctuation parameter $U$ can
be seen in different intermediate states, such as binomial state,
generalized binomial state, hypergeometric state, negative binomial
state, and photon added coherent state. It is also shown that the
depth of nonclassicality can be controlled by various parameters
related to intermediate states. Further, we have provided specific
examples of antibunched states, for which $U$ is greater than its
poissonian state value.
\end{abstract}
\textbf{PACS number(s): 42.50.Lc, 42.50.Ar, 42.50.-p} \vspace{.35cm}


\section{Introduction}

A state which does not have any classical analogue is known as
nonclassical state. For example, squeezed state and antibunched
state are nonclassical. \textcolor{black}{Commonly, standard
deviation of an observable is considered to be the most natural
measure of quantum fluctuation {[}\ref{the:orlowski}] associated
with that observable and the reduction of quantum fluctuation below
the coherent state level corresponds to a nonclassical state. For
example, an electromagnetic field is said to be electrically
squeezed field if uncertainties in the quadrature phase observable
$X$ reduces below the coherent state level (i.e. $\left(\Delta
X\right)^{2}<\frac{1}{2}$) and antibunching is defined as a
phenomenon in which the fluctuations in photon number reduces below
the Poisson level (i.e. $\left(\Delta N\right)^{2}<\langle
N\rangle$) {[}\ref{hbt},\ref{nonclassical}]. Standard deviations can
also be combined to form some complex measures of nonclassicality,
which may increase with the increasing nonclassicality. As an
example, we can note that the total noise} of a quantum state
\textcolor{black}{which,} is a measure of the total fluctuations of
the amplitude\textcolor{black}{, increases with the increasing
nonclassicality in the system {[}\ref{the:orlowski}].} Particular
parameters, which are essentially combination of standard deviations
of some function of quantum phase, were introduced by Carruthers and
Nieto {[}\ref{carutherrs}] as a measure of quantum phase
fluctuations. In recent past people have used Carruthers Nieto
parameters to study quantum phase fluctuations of coherent light
interacting with a nonlinear nonabsorbing medium of inversion
symmetry {[}\ref{Gerry}-\ref{enu:pathak}]. But unfortunately any
discussion regarding the physical meaning of these parameters were
missing since recent past. Recently we have shown that the reduction
of the Carruthers-Nieto symmetric quantum phase fluctuation
parameter $(U)$ with respect to its poissonian state value
corresponds to an antibunched state, but the converse is not true
{[}\ref{the:phase-prakash}]. Consequently reduction of $U$ is a
stronger criterion of nonclassicality than the lower order
antibunching.

The intermediate states are nonclassical in general. It has also been observed that almost all the intermediate
states satisfy the condition of higher order antibunching {[}\ref{A-Verma}]. As the condition of higher order
antibunching is stronger than that of usual antibunching in the sense that a state which is antibunched in the
$lth$ order has to be antibunched in $(l-1)th$ order too but the converse is not true {[}\ref{enu:garcia}].
Therefore, it seems quite reasonable to check whether the intermediate states satisfy the stronger condition of
reduction of $U$ or not. Present study reveals that the intermediate states may satisfy this stronger
nonclassical criterion (i.e. reduction of $U$ criterion).

The introduction of hermitian phase operators have some ambiguities
(interested readers can see the review {[}\ref{Lynch2}]) which lead
to many different formalisms
{[}\ref{enu:L.-Suskind-and}-\ref{enu:bp} and references there in] of
quantum phase. Among the different formalisms, Susskind Glogower
(SG) {[}\ref{enu:L.-Suskind-and}], Pegg Barnett
{[}\ref{enu:D.-T.-Pegg}] and Barnett Pegg (BP) {[}\ref{enu:bp}]
formalisms played most important role in the studies of phase
properties and the phase fluctuations of various physical systems.
For example, SG formalism has been used by Fan {[}\ref{Fan}],
Sanders {[}\ref{Sander}], Yao {[}\ref{Yao}], Gerry {[}\ref{Gerry}],
Carruthers and Nieto {[}\ref{carutherrs}] and many others to study
the phase properties and phase fluctuations. On the other hand Lynch
{[}\ref{Lynch},\ref{Lynch1}], Vacaro {[}\ref{Vacaro}], Tsui
{[}\ref{Y.-K.-Tsui,}], Pathak and Mandal {[}\ref{enu:pathak}] and
others have used the BP formalism for the same purpose. The physical
interpretation of reduction of $U$ is valid in both SG and BP
formalism of quantum phase {[}\ref{the:phase-prakash}]. Here we have
studied the possibilities of observing reduction of $U$ with respect
to its poissonian state value (for intermediate state) in BP
formalism.

\textcolor{black}{The reason behind the study of nonclassical properties of intermediate states lies in the fact
that the most of the interesting recent developments in quantum optics have arisen through the nonclassical
properties of the radiation field only. But the majority of these studies are focused on lowest order
nonclassical effects. Higher order extensions of these nonclassical states, which can be viewed as satisfication
of stronger criterion, have only been introduced in recent past {[}\ref{hong}-\ref{hillery}]. Normally all the
criteria of higher order nonclassicality are stronger than their lower order counter part. As the reduction of
$U$ is also stronger than the usual antibunching criterion. It is of an outstanding curiosity to check whether
this particular stronger criterion of nonclassicality is satisfied by intermediate states or not. }

The importance of a systematic study of quantum phase fluctuation of
intermediate state has also increased with the recent observations
of quantum phase fluctuations in quantum computation
{[}\ref{qutrit}, \ref{L.-L.-Sanchez-Soto}] and superconductivity
{[}\ref{Y.-K.-Tsui,}, \ref{Nature, supercond}] and with the success
in experimental production of photon added coherent state
{[}\ref{the:photonadded-experiment}]. These observations along with
the fact that intermediate states satisfy stronger criterion of
nonclassicality (namely the criterion of HOA ) have motivated us to
study quantum phase fluctuation of intermediate states. In next
section we briefly introduce quantum phase fluctuation parameter
($U$) and the meaning of reduction of $U$. In section 3, it is shown
that the reduction of phase fluctuation parameter $U$ can be seen in
different intermediate state, such as binomial state, hypergeometric
state, generalized binomial state, negative binomial state and
photon added coherent state. Role of various parameters in
controlling the depth of nonclassicality is also discussed. Finally
in section 4 we conclude.

\section{Measures of quantum phase fluctuations: Understanding their physical
meaning }

Dirac {[}\ref{enu:dirac}] introduced the quantum phase operator in 1926. Immediately after Dirac's introductory
work it was realized that the uncertainty relation $\Delta N\Delta\phi\ge\frac{1}{2}$ associated with Dirac's
quantum phase has many problems {[}\ref{Lynch2}]. Later on Louisell {[}\ref{enu:W.-H.-Louisell,}] had shown that
most of the problems can be solved if instead of bare phase operator we consider sine $(S)$ and cosine $(C)$
operators which satisfy \begin{equation}
\begin{array}{c}
[N,C]=-iS\end{array}\label{eq:phase5.1}\end{equation} and \begin{equation}
[N,S]=iC.\label{eq:phase5.2}\end{equation} Therefore, the uncertainty relations associated with them are
\begin{equation} \Delta N\Delta C\ge\frac{1}{2}\left|\langle S\rangle\right|\label{eq:phase5.3}\end{equation}
and \begin{equation} \Delta N\Delta S\ge\frac{1}{2}\left|\langle
C\rangle\right|.\label{eq:phase5.4}\end{equation} There are several formalism of quantum phase, and each
formalism defines sine and cosine in an unique way. The sine and cosine operators in Susskind Glogower formalism
is essentially originated due to a rescaling of the photon annihilation and creation operators with the photon
number operator. Another convenient way is to rescale an appropriate quadrature operator with the averaged
photon number. Barnett and Pegg followed this convention and defined the exponential of phase operator $E$ and
its Hermitian conjugate $E^{\dagger}$ as {[}\ref{enu:bp}]
\begin{equation}
\begin{array}{lcl}
E & = & \left(\overline{N}+\frac{1}{2}\right)^{-1/2}a(t)\\
E^{\dagger} & = & \left(\overline{N}+\frac{1}{2}\right)^{-1/2}a^{\dagger}(t)\end{array}\label{taro}\end{equation}
 where $\overline{N}$ is the average number of photons present in
the radiation field after interaction. The usual cosine and sine of
the phase operator are defined as \begin{equation}
\begin{array}{lcl}
C & = & \frac{1}{2}\left(E+E^{\dagger}\right)\\
S & = &
-\frac{i}{2}\left(E-E^{\dagger}\right)\end{array}\label{chauddo}\end{equation}
which satisfy \begin{equation} \langle C^{2}\rangle+\langle
S^{2}\rangle=1.\label{eq:bp2}\end{equation} Squaring and adding
(\ref{eq:phase5.3}) and (\ref{eq:phase5.4}) we obtain
\begin{equation} (\Delta N)^{2}\left[(\Delta S)^{2}+(\Delta
C)^{2}\right]\left/\left[<S>^{2}+<C>^{2}\right]\right.\geq\frac{1}{4}.\label{eq:babu}\end{equation}
Carruthers and Nieto {[}\ref{carutherrs}] introduced (\ref{eq:babu})
as measure of quantum phase fluctuation and named it as $U$
parameter. To be precise, Carruthers and Nieto defined following
parameter as
a measure of phase fluctuation%
\footnote{They had also introduced two more parameters $S$ and $Q$ for the
purpose of calculation of the phase fluctuations. But these parameters
are not relevant for the present work.%
}: \begin{equation}
U\left(\theta,t,|\alpha|^{2}\right)=(\Delta N)^{2}\left[(\Delta S)^{2}+(\Delta C)^{2}\right]\left/\left[\langle S\rangle^{2}+\langle C\rangle^{2}\right]\right.\label{kuri}\end{equation}
where, $\theta$ is the phase of the input coherent state $|\alpha\rangle$,
$t$ is the interaction time and $|\alpha|^{2}$ is the mean number
of photon prior to the interaction. Later on this parameter draw more
attention and many groups {[}\ref{Gerry}-\ref{enu:pathak}] have
used these parameters as a measure of quantum phase fluctuation.

The total noise of a quantum state is a measure of the total fluctuations
of the amplitude. For a single mode quantum state having density matrix
$\rho$ it is defined as {[}\ref{the:orlowski}]\begin{equation}
\begin{array}{lcl}
T(\rho) & = & (\Delta X)^{2}+(\Delta\dot{X})^{2}\end{array}.\label{eq:total noise}\end{equation}
 In analogy to it we can define the total phase fluctuation as \begin{equation}
T=(\Delta S)^{2}+(\Delta
C)^{2}.\label{eq:totalnoise-phase}\end{equation} Now using the
relations (\ref{eq:phase5.3}), (\ref{eq:phase5.4}), (\ref{eq:bp2}),
(\ref{eq:babu}) and (\ref{eq:totalnoise-phase}) we obtain \[
\left(\Delta N\right)^{2}\left(\Delta S\right)^{2}+\left(\Delta
N\right)^{2}\left(\Delta C\right)^{2}\geq\frac{1}{4}\left(\langle
S\rangle^{2}+\langle C\rangle^{2}\right)=\frac{1}{4}\left(\langle
S^{2}\rangle+\langle C^{2}\rangle-\left(\left(\Delta
S\right)^{2}+\left(\Delta C\right)^{2}\right)\right)\] or,
\[ \frac{1}{4}\left(1-\left(\left(\Delta S\right)^{2}+\left(\Delta
C\right)^{2}\right)\right)\le\left(\left(\Delta S\right)^{2}+\left(\Delta C\right)^{2}\right)\left(\Delta
N\right)^{2}\] or, \begin{equation} U=\frac{\left(\left(\Delta S\right)^{2}+\left(\Delta
C\right)^{2}\right)\left(\Delta N\right)^{2}}{\left(1-\left(\left(\Delta S\right)^{2}+\left(\Delta
C\right)^{2}\right)\right)}=\frac{T\left(\Delta N\right)^{2}}{\left(1-T\right)}\ge\frac{1}{4}.\label{eq:total
noise2}\end{equation} and\begin{equation}
[C,S]=\frac{i}{2}\left(\overline{N}+\frac{1}{2}\right)^{-\frac{1}{2}}.\label{eq:bp3}\end{equation} Therefore,
\begin{equation} (\Delta C)^{2}(\Delta
S)^{2}\geq\frac{1}{16}\frac{1}{\left(\overline{N}+\frac{1}{2}\right)}.\label{eq:bp4}\end{equation}
Now we can write,\[ T=(\Delta C)^{2}+(\Delta S)^{2}\ge(\Delta
C)^{2}+\frac{1}{16\left(\overline{N}+\frac{1}{2}\right)(\Delta
C)^{2}}.\] The function $T=(\Delta
C)^{2}+\frac{1}{16\left(\overline{N}+\frac{1}{2}\right)(\Delta
C)^{2}}$ has a clear minima at $(\Delta
C)^{2}=\frac{1}{4\left(\overline{N}+\frac{1}{2}\right)^{\frac{1}{2}}}$,
which corresponds to a coherent state and thus the total fluctuation
in quantum phase variables $\left((\Delta C)^{2}+(\Delta
S)^{2}\right)$ can not be reduced below its coherent state value
$\frac{1}{2\left(\overline{N}+\frac{1}{2}\right)^{\frac{1}{2}}}$.
Now since $(\Delta N)^{2}$ is positive and the $U=\frac{T(\Delta
N)^{2}}{\left(1-T\right)}=b(\Delta N)^{2}\geq\frac{1}{4}$, therefore
$b=\frac{T}{\left(1-T\right)}$ is positive. Under these conditions
$b$ increases monotonically with the increase in $T$. Thus the
minima of $T$ corresponds to the minima of $b$ too and consequently,
$b$ is minimum for coherent state. In other words $b$ can not be
reduced below its coherent state value. Therefore any reduction in
$U=b(\Delta N)^{2}$ with respect to its poissonian state value will
mean a decrease in $(\Delta N)^{2}$ with respect to its poissonian
state counter part. Thus the reduction of $U$ with respect to its
poissonian state value implies antibunching but the converse is not
true. Earlier we have reported reduction of $U$ with respect to
coherent state in some simple optical processes and verified that
the states are antibunched for the corresponding parameters. But the
fact that every antibunched state is not associated with the
reduction of $U$ was not verified in the earlier work. Here we have
shown that reduction of $U$ is possible for various intermediate
states and have also provided examples of intermediate states, which
are antibunched for specific values of parameters but do not show
reduction of $U$ for the same values of the parameters. The
intermediate states are studied under BP formalism because of the
inherent computational simplicity of these formalism over the
others. As the value of $U$ in coherent (poissonian) state is
$\frac{1}{2}$ our requirement of strong nonclassicality reduces to
\begin{equation} d_{u}=U-\frac{1}{2}<0,\label{eq:du1}\end{equation}
further simplification of the criterion (\ref{eq:du1}) is possible
in BP formalism since the symmetric phase fluctuation parameter in
BP formalism reduces to
$U=[<a^{\mathit{\dagger\mathrm{2}}}a^{2}>+<a^{\mathit{\dagger}}a>-<a^{\mathit{\dagger}}a>^{2}][\frac{<a^{\mathit{\dagger}}a>-<a^{\dagger}><a>+\frac{1}{2}}{<a^{\dagger}><a>}]$
and consequently our requirement for strong nonclassicality is
\begin{equation}
d_{u}=[<a^{\mathit{\dagger\mathrm{2}}}a^{2}>+<a^{\mathit{\dagger}}a>-<a^{\mathit{\dagger}}a>^{2}][\frac{<a^{\mathit{\dagger}}a>-<a^{\dagger}><a>+\frac{1}{2}}{<a^{\dagger}><a>}]-\frac{1}{2}<0.\label{eq:du2}\end{equation}
Now in light of this criterion we would like to study the
nonclassical behavior of intermediate states.

\section{Quantum phase fluctuations in intermediate states}

Actually, an intermediate state is a quantum state which reduces to
two or more distinguishably different states (normally,
distinguishable in terms of photon number distribution) in different
limits. In 1985, such a state was first time introduced by Stoler
\emph{et al.} {[}\ref{stoler}]. To be precise, they introduced
Binomial state (BS) as a state which is intermediate between the
most nonclassical number state $|n\rangle$ and the most classical
coherent state $|\alpha\rangle$. They defined BS as \begin{equation}
\begin{array}{lr}
|p,M\rangle=\sum_{n=0}^{M} & B_{n}^{M}\end{array}|n\rangle=\sum_{n=0}^{M}\sqrt{\left(\begin{array}{c}
M\\
n\end{array}\right)p^{n}(1-p)^{M-n}}|n\rangle\,\,\,\:0\leq p\leq1.\label{eq:binomial1}\end{equation}
This state%
\footnote{The state is named as binomial state because the photon number distribution
associated with this state $\left(i.e.\,|B_{n}^{M}|^{2}\right)$is
simply a binomial distribution.%
} is called intermediate state as it reduces to number state in the
limit $p\rightarrow0$ and $p\rightarrow1$ (as $|0,M\rangle=0$ and
$|1,M\rangle=|M\rangle$) and in the limit of
$M\rightarrow\infty,p\rightarrow1$, where $\alpha$ is a real
constant, it reduces to a coherent state with real amplitude. Since
the introduction of BS as an intermediate state it was always been
of interest to quantum optics, nonlinear optics, atomic physics and
molecular physics community. Consequently, different properties of
binomial state have been studied
{[}\ref{the:hong-chenfu-genralized-BS}-\ref{the:OEBS}]. In these
studies it has been observed that the nonclassical phenomena (such
as, antibunching, squeezing and higher order squeezing) can be seen
in BS. This trend of search for nonclassicality in Binomial state,
continued in nineties and in one hand, several version of
generalized BS has been proposed
{[}\ref{the:hong-chenfu-genralized-BS}-\ref{the:Hong-yi-Fan-generalized-BS}]
and in the other hand people went beyond binomial states and
proposed several other form of intermediate states (such as
\textcolor{black}{odd excited binomial state {[}\ref{the:OEBS}],
hypergeometric state {[}\ref{the:hypergeomtric}], negative
hypergeometric state {[}\ref{the:negativehyper geometric}],
reciprocal binomial state {[}\ref{the:reciprocalBS}], and photon
added coherent state {[}\ref{the:agarwal-photonadded}] etc.).} The
studies in the nineties were mainly limited to theoretical
predictions but the recent developments in the experimental
techniques made it possible to verify some of those theoretical
predictions. For example, we can note that, as early as in 1991
Agarwal and Tara {[}\ref{the:agarwal-photonadded}] introduced photon
added coherent state as \begin{equation}
|\alpha,m\rangle=\frac{a^{\dagger
m}|\alpha\rangle}{\langle\alpha|a^{m}a^{\dagger
m}|\alpha\rangle},\label{eq:photonadded1}\end{equation} (where $m$
is an integer and $|\alpha\rangle$ is coherent state) but the
experimental generation of the state has happened only in recent
past when Zavatta, Viciani and Bellini
{[}\ref{the:photonadded-experiment}] succeed to produce it in 2004.
It is easy to observe that this is an intermediate state, since it
reduces to coherent state in the limit $m\rightarrow0$ and to number
state in the limit $\alpha\rightarrow0$. \textcolor{black}{This
state can be viewed as a coherent state in which additional $m$
photon are added. The photon number distribution of all the above
mentioned states are different but all these states belong to a
common family of states called intermediate state.} It is also been
found that most of these intermediate states show antibunching,
squeezing, higher order squeezing, subpoissonian photon statistics
etc. Inspired by these observations, many schemes to generate
intermediate states have been proposed in recent past
{[}\ref{the:photonadded-experiment},\ref{the:Valverde},\ref{the:RLoFranco1}].
Thus the intermediate states provide a perfect test bed to test the
satisfiability of any new criterion of nonclassicality. Keeping this
in mind we will investigate the possibility of satisfication of
(\ref{eq:du2}) for different intermediate states in the following
subsections.

\subsection{Binomial state}

Binomial state is originally defined as (\ref{eq:binomial1}), from which it is straight forward to show that
\begin{equation}
\begin{array}{lcl}
a|p,M\rangle & = & [Mp]^{\frac{1}{2}}|p,M-1\rangle.\end{array}\label{eq:eigen1}\end{equation}
Similarly, we can write, \begin{equation}
\langle M,p|a^{\dagger}=\langle M-1,p|\left[Mp\right]^{\frac{1}{2}}.\label{eq:bino1}\end{equation}
Consequently, we obtain,

\begin{equation}
\begin{array}{lcl}
\left\langle M,p|a^{\dagger}a|p,M\right\rangle  & = & Mp\end{array},\label{eq:bino2}\end{equation}
\begin{equation}
\left\langle M,p|a^{\dagger2}a^{2}|p,M\right\rangle =M(M-1)p^{2}\label{eq:bino3}\end{equation}
and \begin{equation}
\left\langle a^{\dagger}\right\rangle \left\langle a\right\rangle =Mp(\sum_{n=0}^{M-1}B_{n}^{M-1}B_{n}^{M})^{2}.\label{eq:bino4}\end{equation}
Now using equations (\ref{eq:du2}) and (\ref{eq:eigen1}-\ref{eq:bino4})
one can obtain,

\begin{equation}
\begin{array}{ccc}
d_{U(BS)} & =[\frac{Mp(1-p)}{(\sum_{n=0}^{M-1}B_{n}^{M-1}B_{n}^{M})^{2}} & [\frac{1}{2Mp}+1-(\sum_{n=0}^{M-1}B_{n}^{M-1}B_{n}^{M})^{2}]-\frac{1}{2}]\end{array}\label{eq:du-binomial}\end{equation}

\begin{figure} [h]
\centering \scalebox{0.7}{\includegraphics{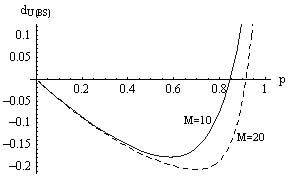}}
\caption{Variation of quantum phase fluctuation of binomial
state.}\label{Fig1}
\end{figure}

From the Fig. 1, it is clear that the binomial state shows reduction
of fluctuation of quantum phase with respect to its coherent state
counter part and thus it satisfies this stronger criterion of
nonclassicality. But it does not satisfy the criterion for higher
values of $p$. In {[}\ref{A-Verma}] we have shown that Binomial
state is always antibunched up to any order. For higher values of
$p$ it is antibunched in every order and thus satisfies the other
strong criterion of nonclassicality but do not satisfy the criterion
laid down on the basis of quantum phase fluctuations. Earlier we had
reported {[}\ref{the:phase-prakash}] that reduction of quantum phase
fluctuation means antibunching but the converse is not true. This is
the first time when an example of such a state which is antibunched
but reduction of quantum phase fluctuation with respect to coherent
state does not happen, is found.

\subsection{Generalized binomial state}

As we have mentioned earlier there are different form of generalized binomial states
{[}\ref{the:hong-chenfu-genralized-BS}-\ref{the:Hong-yi-Fan-generalized-BS}], in the present section we have
chosen generalized binomial state introduced by Roy and Roy {[}\ref{the:broy&proy-generalized-bs}] for our
study. Roy and Roy have introduced the generalized binomial state as \begin{equation}
|N,\alpha,\beta\rangle=\sum_{n=0}^{N}\sqrt{\omega(n,N,\alpha,\beta)}|n\rangle\label{eq:gen-bino1}\end{equation}
where, \begin{equation}
\omega(n,N,\alpha,\beta)=\frac{N!}{(\alpha+\beta+2)_{N}}\frac{(\alpha+1)_{n}(\beta+1)_{N-n}}{n!(N-n)!}\label{eq:gen-bino2}\end{equation}
where $(x)_{r}$ is conventional Pochhammer symbol and $\alpha,\beta>-1$, $n=0,1,....,N$. Now with the help of
properties of Pochhammer symbol and operator algebra we can obtain following relations:\begin{equation}
\begin{array}{lcl}
a|N,\alpha,\beta\rangle & = & \left\{ \frac{N(\alpha+1)}{(\alpha+\beta+2)}\right\} ^{\frac{1}{2}}\sum_{l=0}^{N-1}\left\{ \frac{(N-1)!(\alpha+2)_{l}(\beta+1)_{N-1-l}}{(\alpha+2+\beta+1)_{N-1}l!(N-1-l)!}\right\} ^{\frac{1}{2}}|l\rangle\\
 & = & \left\{ \frac{N(\alpha+1)}{(\alpha+\beta+2)}\right\} ^{\frac{1}{2}}\sum_{n=0}^{N-1}\sqrt{\omega(n,N-1,\alpha+1,\beta)}|n\rangle,\end{array}\label{eq:gb1}\end{equation}
\begin{equation}
\langle N,\alpha,\beta|a^{\dagger}a|N,\alpha,\beta\rangle=\frac{N(\alpha+1)}{(\alpha+\beta+2)},\label{eq:gb2}\end{equation}
\begin{equation}
\langle N,\alpha,\beta|a^{\dagger2}a^{2}|N,\alpha,\beta\rangle=\frac{N(N-1)(\alpha+1)(\alpha+2)}{(\alpha+\beta+2)(\alpha+\beta+3)}\label{eq:gb3}\end{equation}
and \begin{equation}
\begin{array}{ccc}
\left\langle a^{\dagger}\right\rangle \left\langle a\right\rangle  & = & \frac{N^{2}(\alpha+1)^{2}}{(\alpha+\beta+2)^{2}}[\sum_{n=0}^{N-1}\sqrt{\omega(n,N-1,\alpha+1,\beta)}\end{array}]^{2}.\label{eq:gb4}\end{equation}
 Therefore, \begin{equation}
\begin{array}{lcl}
d_{U(GBS)} & = & \left[\frac{(\beta+1)(\alpha+\beta+N+2)}{N^{3}(\alpha+1)(\alpha+\beta+3)[\frac{(N-1)!(\alpha+2)_{n}(\beta+1)_{N-n}}{(\alpha+\beta+3)_{N-1}(N-n-1)!}]}\right.\left[\frac{N(N-1)(\alpha+1)(\alpha+2)}{(\alpha+\beta+2)(\alpha+\beta+3)}\right.\\
 & - & \left.\begin{array}{c}
\left.\frac{N^{4}(\alpha+1)^{2}(N-1)!(\alpha+2)_{n}(\beta+1)_{N-n}}{(\alpha+\beta+2)^{2}(\alpha+\beta+3)_{N-1}(N-n-1)!}+\frac{1}{2}\right]\end{array}-\frac{1}{2}\right]\end{array}\label{eq:gb5}\end{equation}

\begin{figure} [h]
\centering \scalebox{0.7}{\includegraphics{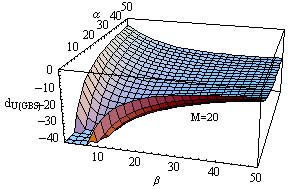}}
\caption{Variation of quantum phase fluctuation with respect to
$\alpha$ and $\beta$ for Roy and Roy generalized binomial
state.}\label{Fig2}
\end{figure}

From Fig. 2 it is clear that the reduction of quantum phase
fluctuation happens for Roy and Roy generalized binomial state. It
is also observed that the depth of nonclassicality reduces with the
increase in $\alpha$ and $\beta$. But as far as the higher (second)
order antibunching is concerned, the depth of nonclassicality
associated with it decreases with increase in $\alpha$ and increases
with increase in $\beta$ (see Fig. 2 and 3 of {[}\ref{A-Verma}]).
Farther it had been observed in {[}\ref{A-Verma}] that for
particular values of $\alpha,\beta$ and $N$ the state does not show
second order antibunching. But here it is found that for the same
values of $\alpha,\beta$ and $N$ we can obtain reduction of quantum
phase fluctuation parameter with respect to its coherent state
counterpart. Consequently we can say that, reduction of quantum
phase fluctuation means antibunching but does not essentially mean
higher order antibunching and therefore, it is not essential that
these two stronger conditions of nonclassicality would appear
simultaneously.

\subsection{Photon added coherent state:}

Photon added coherent state (PACS) is defined as
{[}\ref{the:agarwal-photonadded}])
\begin{equation}
\begin{array}{lcl}
|\alpha,m> & = &
\frac{exp(-|\alpha|^{2}/2)}{[L_{m}(-|\alpha|^{2})m!]^{1/2}}\sum_{n=0}^{\infty}\frac{\alpha^{n}\sqrt{(n+m)!}}{n!}|n+m>\end{array}\label{eq:pacs1}\end{equation}
where $L_{m}(x)$$\begin{array}{cc} = & \sum_{n=0}^{m}\frac{(-x)^{n}m!}{(n!)^{2}(m-n)!}\end{array}$is Lauguere
polynomial. Rigorous operator algebra yields \begin{equation}
<a^{\dagger}a>=\frac{exp(-|\alpha|^{2})}{L_{m}(-|\alpha|^{2})m!}\sum_{n=0}^{\infty}\frac{(n+m)!\alpha^{2(n+1)}(m+n+1)^{2}}{(n+1)!^{2}},\label{eq:pacs2}\end{equation}
\begin{equation}
<a^{\dagger2}a^{2}>=\frac{exp(-|\alpha|^{2})}{L_{m}(-|\alpha|^{2})m!}\sum_{n=0}^{\infty}\frac{(n+m)!\alpha^{n+2}(m+n+1)^{2}(m+n+2)^{2}}{(n+2)!^{2}}\label{eq:pacs3}\end{equation}
and \begin{equation}
<a^{\dagger}>=<a>=\frac{exp(-|\alpha|^{2})}{L_{m}(-|\alpha|^{2})m!}\sum_{n=0}^{\infty}\frac{(n+m)!\alpha^{2n+1}(m+n+1)}{(n+1)(n!)^{2}}.\label{eq:pacs4}\end{equation}
By substituting equations (\ref{eq:pacs2}-\ref{eq:pacs4}) in
(\ref{eq:du2}) we can easily obtain a long expression of $d_{U}$.
Essential characteristic of $d_{U}$ of PACS can be seen in Fig 3. It
is easy to observe that the reduction of quantum phase fluctuation
is possible in photon added coherent state. Since the depth of
nonclassicality increases with the increase in $m$. So we can
conclude, the more photon are added to coherent state the more
nonclassical it is as far as the depth of nonclassicality associated
with quantum phase fluctuation is concerned. This particular
characteristic is also been reflected in higher order antibunching
{[}\ref{A-Verma}]. \textcolor{red}{}

\begin{figure} [h]
\centering \scalebox{0.7}{\includegraphics{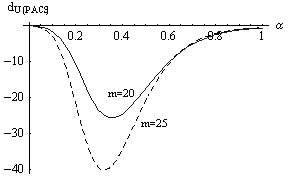}} \caption{Variation of quantum phase fluctuation of photon
added coherent state.}\label{Fig3}
\end{figure}

\subsection{Other intermediate states}

As it is mentioned in the earlier sections, there exist several
different intermediate states. For the systematic study of
possibility of reduction of quantum phase fluctuation in
intermediate states, we have studied all the well known intermediate
states. Since the procedure followed for the study of different
states is similar, mathematical detail has not been shown in the
subsections below. But from the expression of $d_{U}$ and the
corresponding plots it would be easy to see that the reduction of
quantum phase fluctuation can be observed in all the intermediate
states studied below.

\subsubsection{Negative binomial state }

Negative Binomial State(NBS) can be defined as \begin{equation}
|p,M\mbox{>}=\sum_{n=M}^{\infty}[\left(\begin{array}{c}
n\\
M\end{array}\right)p^{M+1}(1-p)^{n-M}]^{1/2}|n>.\label{eq:nbs}\end{equation}
Similar operator algebra yields

\begin{equation}
d_{U(NBS)}=\left[\frac{(M+1)(1-p)^{2M+1}}{p^{2(M+2)}}\left\{ \frac{\frac{(M+1-p)}{p}-\frac{p^{2(M+1)}}{(1-p)^{2M}}\left(\sum_{n=M}^{\infty}\sqrt{\left(\begin{array}{c}
n+1\\
M\end{array}\right)\left(\begin{array}{c}
n\\
M\end{array}\right)(1-p)^{(n+1)}(n+1)}\right)^{2}+\frac{1}{2}}{\left(\sum_{n=M}^{\infty}\sqrt{\left(\begin{array}{c}
n+1\\
M\end{array}\right)\left(\begin{array}{c}
n\\
M\end{array}\right)(1-p)^{(n+1)}(n+1)}\right)^{2}}\right\}
-\frac{1}{2}\right].\label{eq:du-2}\end{equation} Variation of
quantum phase fluctuation parameter with respect to the probability
$p$ for NBS is shown in the Fig. 4 and it has been observed that the
state is more nonclassical for lower values of $p$ and higher values
of $M$ {[}\ref{A-Verma}]. This is consistent with the earlier
observations on higher order antibunching of NBS.

\begin{figure} [h]
\centering \scalebox{0.7}{\includegraphics{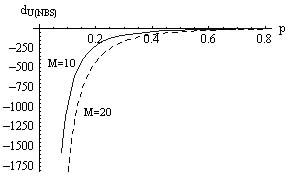}}
 \caption{Variation of quantum phase fluctuation of
negative binomial state.}\label{Fig4}
\end{figure}

\subsection{Hyper geometric state}

Hyper geometric state (HS) is defined as {[}\ref{the:hypergeomtric}]
\begin{equation}
|L,M,p>=\sum_{n=0}^{M}H_{n}^{M}(p,L)|n>\label{eq:hs1}\end{equation}
where, $H_{n}^{M}=\left[\left(\begin{array}{c}
Lp\\
n\end{array}\right)\left(\begin{array}{c}
L(1-p)\\
M-n\end{array}\right)\right]^{1/2}\left(\begin{array}{c}
L\\
M\end{array}\right)^{-1/2}$. For this intermediate state we obtain \begin{equation}
\begin{array}{lcl}
d_{U(HS)} & = & \left[\frac{pM(1-p)(L-M)\left(\begin{array}{c}
L\\
M\end{array}\right)^{2}}{(L-1)\sqrt{Lp}\left(\sum_{n=0}^{M-1}\sqrt{\left(\begin{array}{c}
Lp\\
n\end{array}\right)\left(\begin{array}{c}
L(1-p)\\
M-n\end{array}\right)\left(\begin{array}{c}
Lp-1\\
n\end{array}\right)\left(\begin{array}{c}
L(1-p)\\
M-n-1\end{array}\right)}\right)^{2}}\right.\\
 & \times & \left.\left\{ Mp-\frac{\sqrt{Lp}}{\left(\begin{array}{c}
L\\
M\end{array}\right)^{2}}\left(\sum_{n=0}^{M-1}\sqrt{\left(\begin{array}{c}
Lp\\
n\end{array}\right)\left(\begin{array}{c}
L(1-p)\\
M-n\end{array}\right)\left(\begin{array}{c}
Lp-1\\
n\end{array}\right)\left(\begin{array}{c}
L(1-p)\\
M-n-1\end{array}\right)}\right)^{2}+\frac{1}{2}\right\} -\frac{1}{2}\right].\end{array}\label{eq:du-hs}\end{equation}

\begin{figure} [h]
\centering \scalebox{0.7}{\includegraphics{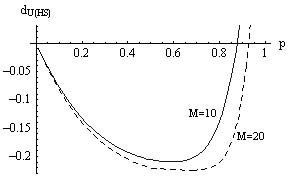}} \caption{Variation of quantum phase fluctuation of hyper
geometric state.}\label{Fig5}
\end{figure}

Fig. 5 depicts the characteristics of quantum phase fluctuation of
HS. From this figure we can clearly observe that HS does not satisfy
the reduction of quantum phase fluctuation criterion for higher
values of $p$ but it satisfies the condition of higher order
antibunching and consequently the condition of antibunching for
those values of $p.$ This is consistent with our theoretical
prediction that reduction of quantum phase fluctuation means
antibunching but the converse is not true.

\section{Conclusions}

In essence, all the intermediate states described above \textcolor{black}{show
reduction of $U$ with respect to its coherent state value}%
\footnote{In reciprocal binomial state, we have not observed this phenomenon. \textcolor{red}{}%
}\textcolor{black}{. This establishes that the intermediate states
can satisfy the stronger criterion of nonclassicality compared to
the criterion of usual antibunched state. This is consistent with
our earlier observation of higher order antibunching of intermediate
states {[}\ref{A-Verma}]. Further, we would like to note} that the
binomial state can show reduction of fluctuation of quantum phase
with respect to its coherent state counter part and thus it can
satisfy this stronger criterion of nonclassicality. But it does not
satisfy the criterion for higher values of $p$ (see Fig. 1). In
{[}\ref{A-Verma}] we have shown that Binomial state is always
antibunched up to any order. Thus for higher values of $p$ it is
higher order antibunched and consequently satisfies the other strong
criterion of nonclassicality but do not satisfy the criterion laid
down on the basis of quantum phase fluctuations. Similar phenomenon
is also observed in HS. Earlier we had reported
{[}\ref{the:phase-prakash}] that reduction of quantum phase
fluctuation means antibunching but the converse is not true. This is
the first time when an example of such a state which is antibunched
but does not show reduction of quantum phase fluctuation with
respect to coherent state, is found. Further from the study of phase
properties of Roy Roy GBS and HS we have learnt that the reduction
of quantum phase fluctuation mean antibunching but does not
essentially mean higher order antibunching and therefore, it is not
essential that these two stronger conditions of nonclassicality
appear simultaneously. In connection to PACS we have observed that
the more photon are added to coherent state the more nonclassical
the PACS, is as far as the depth of nonclassicality associated with
quantum phase fluctuation is concerned. This particular
characteristic has also been reflected in higher order antibunching
{[}\ref{A-Verma}]. \textcolor{red}{}Further we have seen that the
NBS is more nonclassical for lower values of $p$. This is also
similar with the earlier observations on higher order antibunching
of NBS.

\textbf{Acknowledgement}: AP thanks to DST, India for partial financial
support through the project grant SR\textbackslash{}FTP\textbackslash{}PS-13\textbackslash{}2004.

\vspace{.7cm}
 \textbf{References}

\begin{enumerate}
\item \label{the:orlowski}Orlowski A, \emph{Phys. Rev. A} 48 (1993) 727.
\item \label{hbt}Hanbury-Brown R, Twiss R Q, \emph{Nature} 177 (1956) 27.
\item \label{nonclassical}Dodonov V V, J\emph{. Opt. B. Quant. and Semiclass.
Opt.} 4 (2002) R1.
\item \label{carutherrs} P. Carutherrs and M. M. Nieto, \emph{Rev. Mod.
Phys.} \textbf{40} (1968) 411.
\item \label{Gerry}C. C. Gerry, \emph{Opt. Commun.,} \textbf{63} (1987)
278.
\item \label{the:gerry1}C. C. Gerry, \emph{Opt. Commun.,} \textbf{75} (1990)
168.
\item \label{Lynch}R. Lynch, \emph{Opt. Commun.,} \textbf{67} (1988) 67.
\item \label{enu:pathak}A. Pathak and S. Mandal, \emph{Phys. Lett. A} 272
(2000) 346.
\item \label{the:phase-prakash}P Gupta and A Pathak, \emph{Phys. Lett.
A} 365 (2007) 393.
\item \label{A-Verma}A Verma, N K Sharma and A Pathak, quant-ph\textbackslash{}0706.0697.
\item \label{enu:garcia}A. Pathak and M. E. Garcia, \emph{Appl. Phys. B}
\textbf{84} (2006) 479.
\item \label{Lynch2} R. Lynch, \emph{Phys. Reports} \textbf{256} (1995)
367.
\item \label{enu:L.-Suskind-and}L. Suskind and J. Glogower, \emph{Physics}
\textbf{1} (1964) 49.
\item \label{enu:D.-T.-Pegg}D. T. Pegg and S. M. Barnett, \emph{Phys. Rev.
A} \textbf{39} (1989) 1665.
\item \label{enu:bp} S. M. Barnett and D. T. Pegg, \emph{J. Phys. A} \textbf{19}
(1986) 3849.
\item \label{Fan}Fan Hong-Yi and H. R. Zaidi, \emph{Opt. Commun.,} \textbf{68}
(1988) 143.
\item \label{Sander}B. C. Sanders, S. M. Barnett and P. L. Knight, \emph{Opt.
Commun.}, \textbf{58} (1986) 290.
\item \label{Yao}D. Yao, \emph{Phys. Lett. A}, \textbf{122} (1987) 77.

\item \label{Lynch1}R. Lynch, \emph{J. Opt. Soc.Am,} \textbf{B4} (1987)
1723.
\item \label{Vacaro}J. A. Vaccaro and D. T. Pegg, \emph{Opt.Commun.,} \textbf{70}
(1989) 529.
\item \label{Y.-K.-Tsui,}Y. K. Tsui, \emph{Phys Rev. A} \textbf{47} 12296
(1993).
\item \textcolor{black}{\label{hong}Hong C K and Mandel L,} \textcolor{black}{\emph{Phys.
Rev. Lett.}} \textbf{\textcolor{black}{54}} \textcolor{black}{(1985)
323.}
\item \textcolor{black}{\label{lee1}Lee C T,} \textcolor{black}{\emph{Phys.
Rev. A}} \textbf{\textcolor{black}{41}} \textcolor{black}{(1990) 1721.}
\item \textcolor{black}{\label{hillery}M. Hillery,} \textcolor{black}{\emph{Phys.
Rev. A}} \textbf{\textcolor{black}{36}} \textcolor{black}{(1987) 3796.}
\item \label{qutrit}A B Klimov \emph{et al, J. Phys. A} \textbf{37} (2004)
4097.
\item \label{L.-L.-Sanchez-Soto}L. L. Sanchez-Soto \emph{et al Phys. Rev.
A} \textbf{66} (2002) 042112.
\item \label{Nature, supercond}I. Iguchi, T. Yamaguchi and A. Sugimato
\emph{Nature} \textbf{412} (2001) 420 .
\item \label{the:photonadded-experiment}A Zavatta, S Viciani, M Bellini,
\emph{Science} \textbf{306} (2004) 660.
\item \label{enu:dirac}P. A . M. Dirac, \emph{Proc. Royal. Soc. London}
\textbf{Ser. A 114} (1927) 243.
\item \label{enu:W.-H.-Louisell,}W. H. Louisell, \emph{Phys. Lett.} \textbf{7}
(1963) 60.
\item \label{stoler}D Stoler, B E A Saleh and M C Teich, \emph{Opt. Acta}
\textbf{32} (1985) 345.
\item \label{the:hong-chenfu-genralized-BS}Hong-Chen Fu and Ryu Sasaki
\emph{J. Phys. A} \textbf{29} (1996) 5637.
\item \label{the:broy&proy-generalized-bs}P Roy and B Roy, \emph{J. Phys.
A} \textbf{30} (1997) L719.
\item \label{the:Hong-yi-Fan-generalized-BS}Hong-Yi Fan and Nai-le Liu,
\emph{Phys. Lett. A} \textbf{264} (1999) 154.
\item \label{the:OEBS}A S F Obada, M Darwish and H H Salah, I\emph{nt.
J. Theo. Phys.} \textbf{41} (2002) 1755.
\item \label{the:hypergeomtric}H. C. Fu and R. Sasaki, \emph{J. Math. Phys.}
38 (1997) 2154.
\item \label{the:negativehyper geometric}H Fan and N Liu, \emph{Phys. Lett.
A} \textbf{250} (1998) 88.
\item \label{the:reciprocalBS}M H Y Moussa and B Baseia, \emph{Phys. Lett.
A} \textbf{238} (1998) 223.
\item \label{the:agarwal-photonadded}G S Agarwal and K Tara, \emph{Phys.
Rev. A} \textbf{43} (1991) 492.
\item \label{the:Valverde}C Valverde \emph{et al}, \emph{Phys. Lett. A}
\textbf{315} (2003) 213.
\item \label{the:RLoFranco1}R Lo Franco \emph{et al}, \emph{Phys. Rev.
A} \textbf{74} (2006) 045803.
\end{enumerate}

\end{document}